\documentclass{pkas}

\def\year{2014} 
\def\month{May} 
\def\volume{00} 
\def\issue{0} 
\def\beginpage{1} 
\def\endpage{99} 
\def\received{February 30, 2014} 
\def\accepted{February 31, 2014} 
\date{Received \received ; accepted \accepted}

\title{
Strong gravitational lenses and multi-wavelength galaxy surveys with AKARI, Herschel, SPICA and Euclid
}

\author[1]{Stephen~Serjeant}

\affil[1]{Department of Physical Sciences, The Open University, Milton Keynes, MK7 6AA, UK}

\def\corrauthor{
S. Serjeant
}

\def\runningauthor{
S. Serjeant
}

\def\runningtitle{
Strong gravitational lenses \& multi-wavelength galaxy surveys
}

\def\keywords{
cosmology: observations ---
galaxies: evolution ---
galaxies: formation ---
galaxies: starburst ---
infrared: galaxies ---
submillimeter: galaxies
}

\def\abstracttext{
Submillimetre and millimetre-wave surveys with {\it Herschel} and the South Pole Telescope have revolutionised the discovery of strong gravitational lenses. Their follow-ups have been greatly facilitated by the multi-wavelength supplementary data in the survey fields. The forthcoming {\it Euclid} optical/near-infrared space telescope will also detect strong gravitational lenses in large numbers, and orbital constraints are likely to require placing its deep survey at the North Ecliptic Pole (the natural deep field for a wide class of ground-based and space-based observatories including {\it AKARI}, {\it JWST} and {\it SPICA}).  In this paper I review the current status of the multi-wavelength survey coverage in the NEP, and discuss the prospects for the detection of strong gravitational lenses in forthcoming or proposed facilities such as {\it Euclid}, {\it FIRSPEX} and {\it SPICA}.
}

\begin{document}
\pkashead 


\section{Introduction\label{sec:intro}}

\subsection{Strong gravitational lensing surveys}\label{sec:lensingintro}
There are about a couple of hundred or so known strong galaxy-galaxy gravitational lenses discovered through optical/near-infrared and radio surveys. Negrello et al. (2010) demonstrated a dramatically efficient alternative method of lens discovery, using the strong magnification bias present in the steep bright-end slope of submm-wave number counts. This method, and its generalisation to the bright end of the submm luminosity function, promise to deliver a further $\sim1000$ lensing systems from {\it Herschel} and South Pole Telescope wide-field surveys. Many cosmological applications of lensing are limited by sample size (e.g. Treu 2010; see also Serjeant 2014).  This paper briefly reviews some prospects for further lens discovery and follow-ups in deep-field surveys, wide-field surveys and pointed observations with future optical, infrared and radio facilities.

\subsection{The North Ecliptic Pole surveys\label{sec:nep}}
The North Ecliptic Pole (NEP) is the natural deep field for space telescopes on, for example, approximately polar orbits around the Earth or L2. For example, the NEP is in the continuous viewing zone for the {\it Hubble Space Telescope}, the {\it James Webb Space Telescope}, {\it AKARI} and {\it SPICA} among others. It is almost inevitably a site of the deep calibration fields for {\it Euclid}. Section \ref{sec:results} below will present predictions for the numbers of high redshift strongly gravitationally lensed {\it Euclid} galaxies, both in its deep 40\,deg$^2$ survey (e.g. the NEP) and its wide $15\,000$\,deg$^2$ survey. 

\begin{figure}
\centering
\vspace*{-3cm}
\hspace*{-1cm}
\includegraphics[width=120mm]{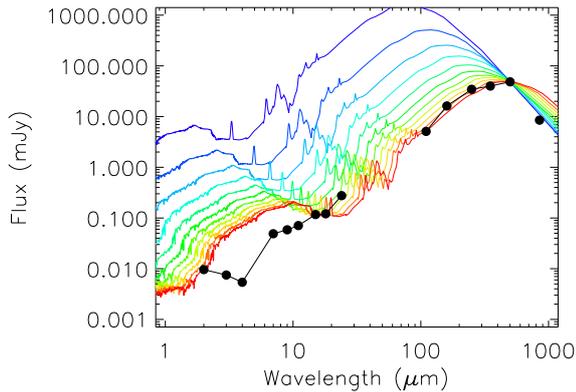}
\vspace*{-8cm}
\caption{An M82 spectral energy distribution, redshifted from $z=0$ in steps of $\Delta z=0.5$, normalised to the NEP depth at $500\,\mu$m. Also shown are the $5\sigma$ depths from the {\it AKARI} NEP-deep field (near/mid-infrared), the {\it Herschel} SPIRE data ($250-500\,\mu$m) and the SCUBA-2 Cosmology Legacy Survey data ($850\,\mu$m), and the $3\sigma$ {\it Herschel} PACS data ($110-160\,\mu$m).\label{fig:m82}}
\end{figure}

The NEP is already the site of the premier {\it AKARI} extragalactic survey field (e.g. Matsuhara et al. 2006), covering $\sim0.5$\,deg$^2$ in the NEP-Deep survey and $5.6\,$deg$^2$ in the NEP-Wide survey. This entire field has been covered by a {\it Herschel} SPIRE $9\,$deg$^2$ map at $250-500\,\mu$m (Serjeant et al. in prep.), including deeper pointings in the deep survey field, and deep PACS map covering the majority of the {\it AKARI} NEP-Deep survey field. The NEP has been observed by WSRT and GMRT, and is also one of the deep cosmological survey fields of the Low Frequency Array (LOFAR). Although inaccessible to ALMA, the field is accessible to IRAM and its forthcoming upgrade NOEMA, and at the time of writing the NEP-Deep field has just been mapped in a PONG 2700 arcsecond diameter SCUBA-2 map at $850\,\mu$m. With the majority ownership of the James Clerk Maxwell Telescope telescope shortly passing to the East Asian partnership, the prospects may be good for extending this submm coverage. Figure~\ref{fig:m82} compares the {\it AKARI}, {\it Herschel} and {\it SCUBA-2} depths to redshifted M82 template spectral energy distributions.

\section{Gravitational lensing methodology}\label{sec:method}
The success of the Negrello et al. (2010) methodology for finding gravitational lenses makes it attractive to search for other applications of magnification bias. For example, in the nearby Universe, galaxies with highest infrared luminosities have a smaller proportion of their luminosity in the [C{\sc ii}] $158\,\mu$m line. The  [C{\sc ii}] luminosity function should therefore have a steeper bright-end slope than that of the far-infrared luminosity function. However, the ``deficit'' in [C{\sc ii}] and other far-infrared lines seen in low redshift starbursts is not seen consistently at higher redshifts (e.g. Rigopoulou et al. 2014), so mapping the sky for redshifted far-infrared lines with e.g. {\it SPICA} or {\it FIRSPEX} will not necessarily be an efficient lens selection in itself  -- at least, not obviously more effective than far-infrared luminosities. 

However there are several other galaxy distribution functions with exponential cut-offs, such as the bright end of the Ly$\alpha$ luminosity function, the bright end of the H$\alpha$ luminosity function, and the high mass end of the H{\sc i} mass function. As the star formation rate of galaxies increases, there is an increasing fraction of the ionizing radiation from the starburst intercepted by dust and re-radiated as thermal radiation. Therefore, extreme H$\alpha$-bright and Ly$\alpha$-bright galaxies should be rarer than their far-infrared counterparts. 

To predict the numbers of lensing systems, I assume the probability of lensing magnification $\mu$ at redshift $z$ is $p(\mu,z)=a(z)\mu^{-3}$, where $a(z)$ follows the predictions of Perrotta et al. (2002, 2003). This formalism assumes singular isothermal sphere (SIS) lenses, and a mass spectrum following Sheth \& Tormen (1999). I assume two cases maximum lens magnification, 10 and 30, equivalent to source physical sizes of $\sim1-10h^{-1}$\,kpc. I also present predictions for a non-evolving SIS lens population with the lensing optical depth $a$ normalised to the Perrotta predictions at $z=1$. 

\begin{figure*}[t]
\centering
\includegraphics[angle=0,width=170mm]{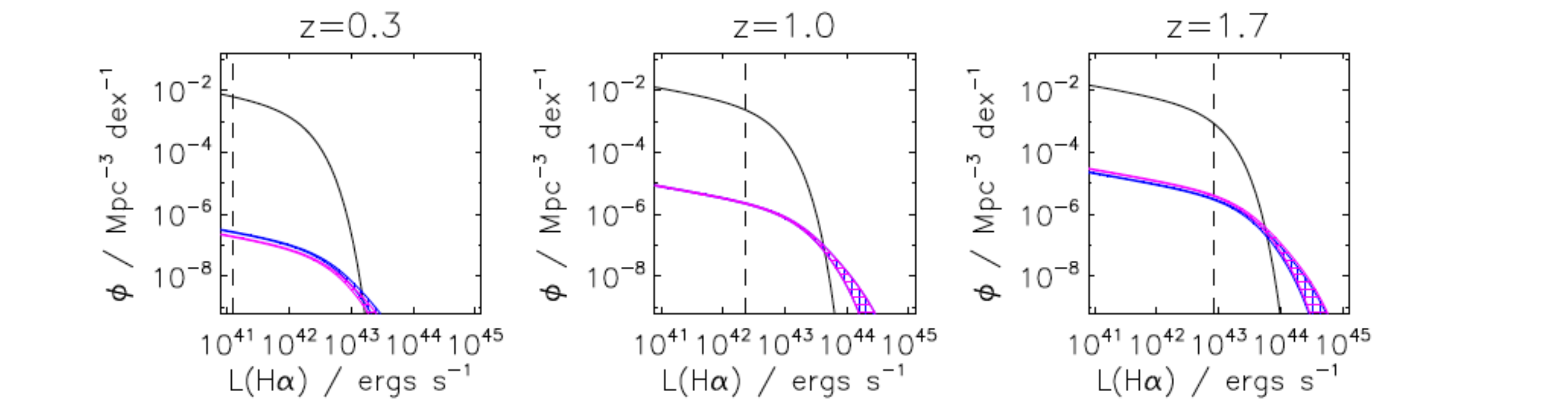}
\includegraphics[angle=0,width=170mm]{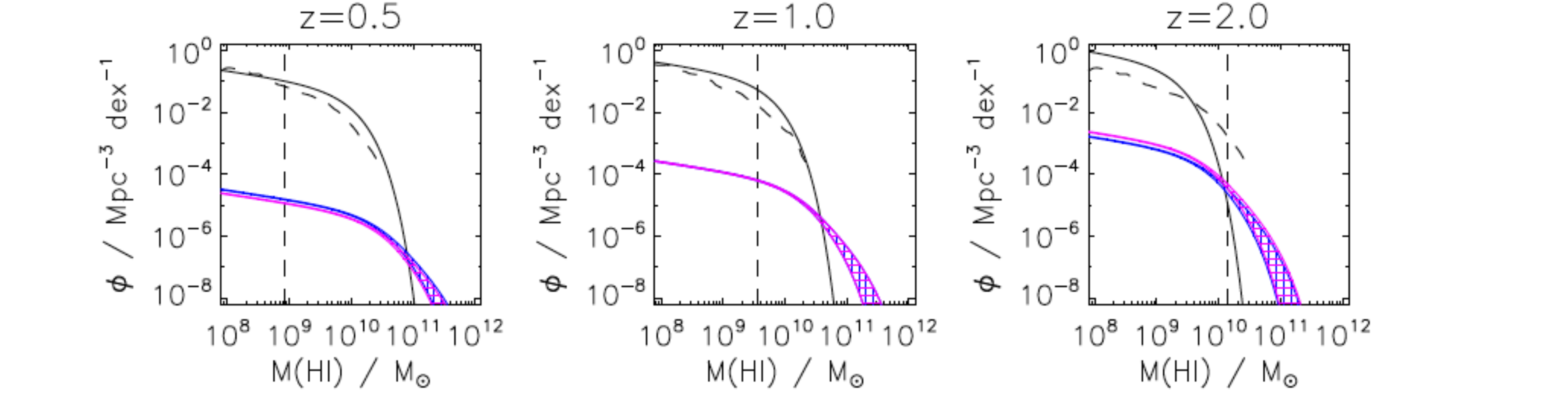}
\includegraphics[angle=0,width=170mm]{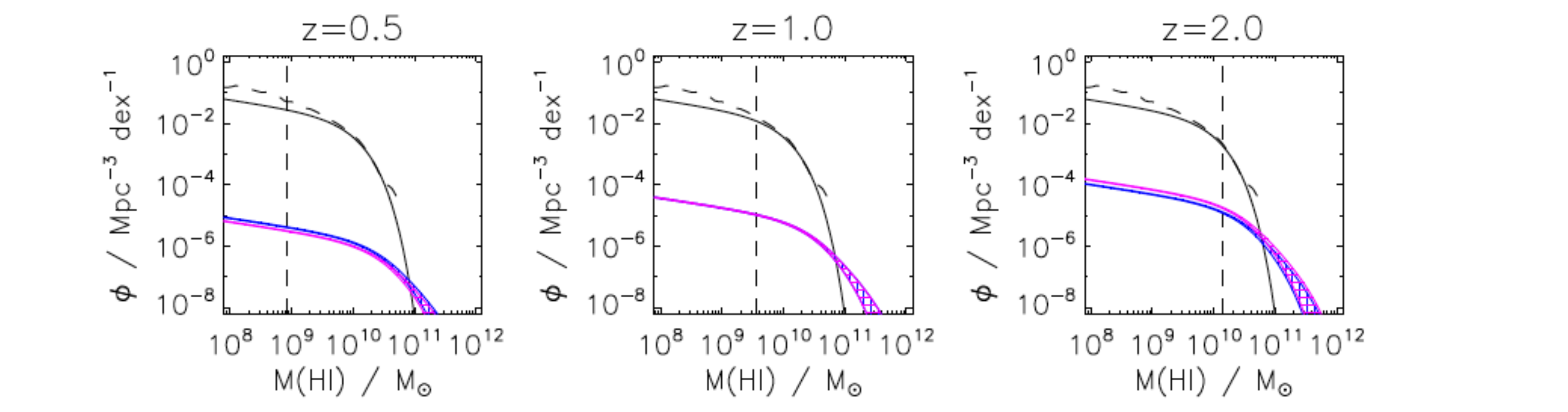}
\caption{Predicted gravitationally lensed galaxies with maximum magnifications $10-30$, for the no-evolution lens model (dark blue) and the Perrotta (2002, 2003) model (pink), for background sources at various redshifts. The top panels show the H$\alpha$ luminosity function of Geach et al. (2010), while the central panels show the analytic H{\sc i} mass function predicted by Abdalla et al. (2010, full line) and the semi-analytic predicted H{\sc i} mass function of Lagos et al. (2012, dashed). The vertical dashed lines show the 5$\sigma$ sensitivity limit of future dark energy experiments: {\it Euclid} wide (top) and the proposed full {\it SKA} dark energy survey of Abdalla et al. (2010) (centre and bottom). The bottom row shows the most conservative {\it SKA} prediction, the unevolving H{\sc i} mass function case. Figure partly adapted from Serjeant (2014).\label{fig:lfs}}
\end{figure*}

\begin{figure*}[t]
\centering
\vspace*{-4cm}
\includegraphics[width=170mm]{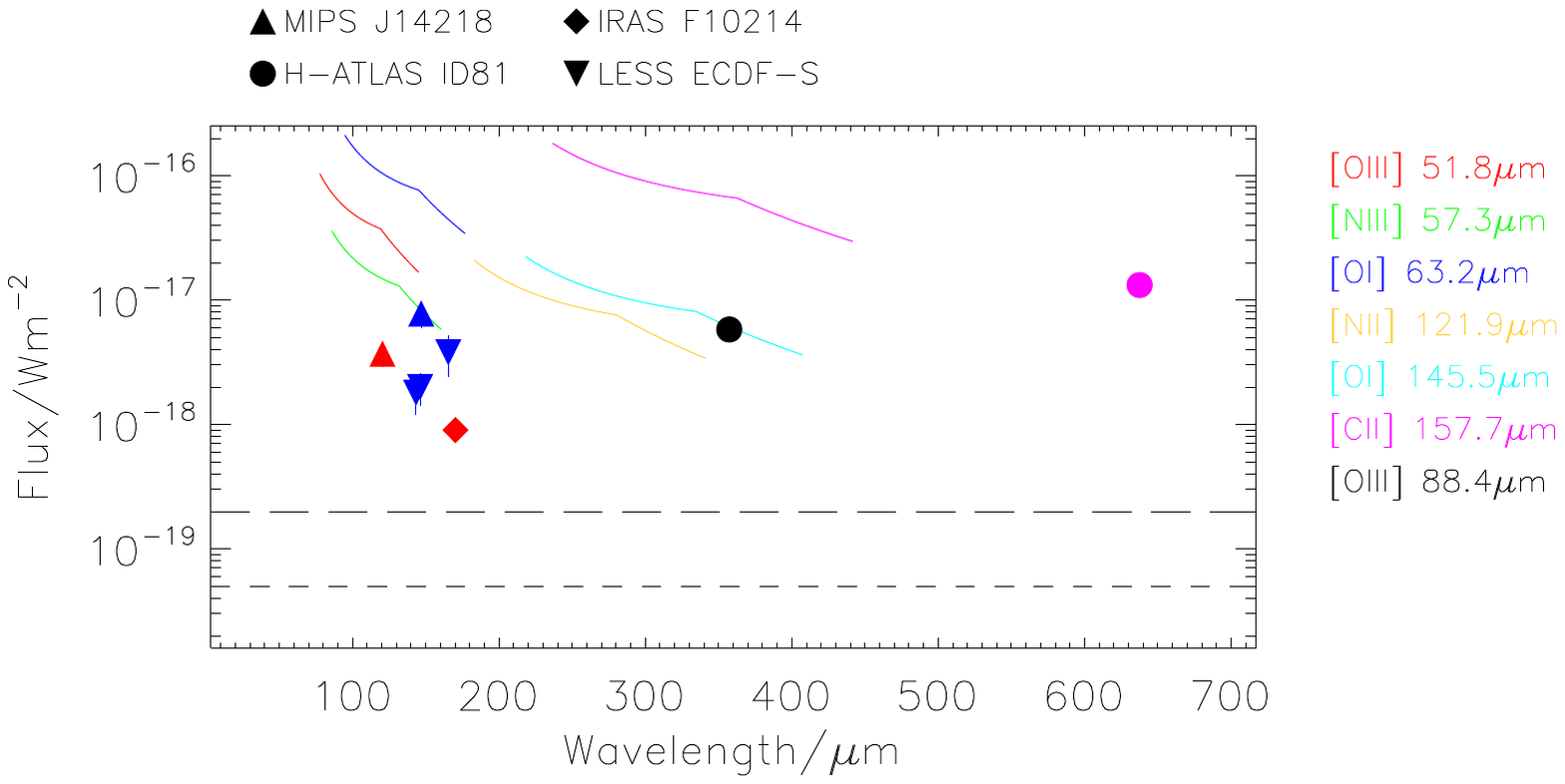}
\vspace*{-11.5cm}
\caption{Predicted line fluxes for the {\it Euclid}-wide H$\alpha$ gravitational lenses. The required and goal $5\sigma$ 1 hour sensitivities of SAFARI are shown as horizontal dashed lines (extending through and beyond the SAFARI wavelength range). Also shown are the observed fluxes for a variety of unlensed deep-field submm galaxies and lensed infrared-selected galaxies. The symbol shape indicates the sample, while the colours identify the lines annotated to the right. \label{fig:spica}}
\end{figure*}

\section{Results: predictions for Euclid and SKA}\label{sec:results}

Figure~\ref{fig:lfs} shows the predictions for the lensed H$\alpha$-emitting population. Using the Geach et al. (2010) luminosity function and a $>12L_*$ selection for strong lens candidates, {\it Euclid}-wide will detect $\sim10^3$ gravitational lenses with $\sim97$\%\ reliability. While only $\sim1$\% of the potential total number of strong lensing events in {\it Euclid}, the reliability of morphologically-selected lenses is much lower, and the H$\alpha$-selected lensing systems have the further advantage of prior knowledge of the source redshift, and (for early-type lenses) fundamental plane constraints on the foreground lens redshifts. 

Not shown in Figure~\ref{fig:lfs} are the predictions for gravitationally lensed Ly$\alpha$-emitters in the 40\,deg$^2$  {\it Euclid}-deep field (e.g. NEP). Using the Ly$\alpha$ luminosity function of Ouchi et al. (2010), I predict between $67-311$ lensed galaxies at $z>6$, $13-74$ at $z>7$ and $2-16$ at $z>8$. This approximately doubles the numbers of high redshift galaxies seen in {\it Euclid}-deep. As with the unlensed high-redshift population, they will need to be separated from the lower-redshift H$\alpha$ line-emitting population. This will be investigated in more detail in Marchetti \& Serjeant (in prep.). A similar effect has also been predicted for ultra-high-redshift galaxies selected in the rest-frame ultraviolet continuum with the {\it James Webb Space Telescope} (Wyithe et al. 2011). 

Even larger numbers of strong gravitational lenses are possible with the proposed dark energy experiment with the 1-year full  {\it SKA}  (Abdalla et al. 2010). In the conservative no-evolution case, $\sim10^4$ strong gravitational lenses are predicted at $>12M_*$, again with a similarly extraordinarily high reliability. For the evolving mass function of Abdalla et al. (2010), this increases to an astonishing $\sim10^5$ (Serjeant 2014). The semi-analytic predictions of Lagos et al. (2011) lie between these two cases. 

\section{Discussion: prospects for FIRSPEX and SPICA}\label{sec:discussion}

The H$\alpha$-selected lensed systems will be among the brightest high-redshift line emitters on the sky, so will be unchallenging targets for the E-ELT in H$\alpha$ and rest-frame optical continuum. Even in Paschen $\alpha$, a 1-hour E-ELT observation would be sufficient to resolve the kinematics with the METIS instrument. The gravitationally lensed Ly$\alpha$ populations will be among the brightest known line emitters at $z>6$ and excellent targets for TMT (in the North) and E-ELT (in the South). 

A back of the envelope calculation easily demonstrates excellent prospects for the detectability of far-infrared lines in at least some H$\alpha$-selected gravitational lenses. 
The typical line far-infrared fluxes for $z\sim0.2$ ultraluminous infrared galaxies are $\sim(1-10)\times10^{-17}$\,W\,m$^{-2}$ (Farrah et al. 2013), while the typical H$\alpha$ line fluxes for an approximately identical sample are $\sim1\times10^{-17}$\,W\,m$^{-2}$ (Veilleux, Kim \& Sanders 1999), i.e. between about 10\% and 100\% of the far-infrared fluxes. The H$\alpha$ flux limit for the {\it Euclid} lens sample at $z=0.5-1.8$ is $(3.6-22)\times10^{-18}$\,W\,m$^{-2}$, while the {\it SPICA} SAFARI $5\sigma$ $1$\,hour sensitivity (goal) is $2\times10^{-19}$\,W\,m$^{-2}$ ($1\times10^{-20}$\,W\,m$^{-2}$). 

Figure~\ref{fig:spica} shows a scaling from the H$\alpha$ line fluxes (Veilleux et al. 1999) to the far-infrared lines in the $z\sim0.2$ ultraluminous galaxy sample. The predicted line fluxes are far above the {\it SPICA} SAFARI instrument limits. It does not follow that all the {\it Euclid} H$\alpha$ lenses will be this bright, since a {\it less}-obscured starburst will have a lower far-infrared continuum and line flux. Nonetheless, the increasing predominance of ultraluminous starbursts at high redshift (e.g. Le Floc'h et al. 2005) makes it plausible that at least $\sim10\%$ will be this bright. 

It is also instructive to compare these predictions with observations of strongly lensed submm galaxies. Figure~\ref{fig:spica} shows that the observations of H-ATLAS ID81 are comparably bright (Valtchanov et al. 2011), though at higher redshift. This is perhaps the most obvious comparator to the {\it Euclid} sample, since it is selected in a wide-field survey. Rather fainter are the LESS sample (Coppin et al. 2012), though as these are from a pencil-beam submm survey in the extended Chandra Deep Field South, this is perhaps to be expected. More surprising though are the faint far-infrared line fluxes of MIPS J142824.0+352619 and IRAS F10214+4724 (Sturm et al. 2010). The latter would just pass the {\it Euclid} H$\alpha$ lens selection flux (Serjeant et al. 1998), though it just exceeds the high redshift limit. Even so, these fluxes are still far above the SAFARI detection limit. 

This also implies that the proposed {\it FIRSPEX} mission, which plans to survey the whole sky in key far-infrared emission lines, may also be able to detect strongly gravitationally-lensed high-redshift galaxies. From Figure~\ref{fig:spica}, one can see that a [C{\sc i}] $370\,\mu$m {\it FIRSPEX} window at a sensitivity of $\sim5\times10^{-17}$\,W\,m$^{-2}$ (5\,GHz instantaneous bandwidth) will be able to detect serendipitous redshifted [C{\sc ii}] $158\,\mu$m emission in a redshift interval of $\Delta z=0.014$, i.e. up to $\sim15$ H$\alpha$-selected lensed galaxies in the 15\,000\,deg$^2$ {\it Euclid}-wide survey, and proportionately more strong gravitational lenses in the proposed {\it FIRSPEX} all-sky survey.

The far-infrared lines will be resolved spectrally but not spatially. Differential magnification may affect diagnostic line ratios, which may be addressed by modelling of single systems or of the population (e.g. Serjeant 2012). Alternatively, the similarity of emission line profiles with those of spatially-resolved lines can be used to argue that their magnifications are approximately equal (e.g. Omont et al. 2013).

\section{Conclusions}\label{sec:conclusions}


{\it Euclid} will be able to provide a $\sim100\%$ reliable sample of $\sim10^3$ strong gravitational lens events with known background redshifts by selecting $>12L_*$ H$\alpha$-emitting galaxies at redshifts $z<1.7$. These objects will be unchallenging targets for {\it SPICA}, TMT and E-ELT, and some may even be detectable by {\it FIRSPEX}. The numbers of $z>6$ galaxies in {\it Euclid}'s deep survey (e.g. in the NEP, also a deep AKARI/{\it Herschel}/LOFAR/SCUBA-2 field) will be approximately doubled by strong galaxy-galaxy lensing. The proposed 1-year dark energy survey with the full SKA will generate up to $10^5$ reliable lensing events with known source redshifts, via $>12M_*$ H{\sc i} selection at $z<2$.

\acknowledgments

The author would like to thank many colleagues in the {\it Eucid} Strong Lensing and Galaxy Evolution working groups for stimulating discussions. This work was supported by the Science and Technology Facilities Council under grant ST/J001597/1.

%


\end{document}